# Yacc is dead

Matthew Might and David Darais

University of Utah, Salt Lake City, Utah, USA,
`might@cs.utah.edu`, `darais@cs.utah.edu`

**Abstract.** We present two novel approaches to parsing context-free languages. The first approach is based on an extension of Brzozowski's derivative from regular expressions to context-free grammars. The second approach is based on a generalization of the derivative to parser combinators. The payoff of these techniques is a small (less than 250 lines of code), easy-to-implement parsing library capable of parsing arbitrary context-free grammars into lazy parse forests. Implementations for both Scala and Haskell are provided. Preliminary experiments with S-Expressions parsed millions of tokens per second, which suggests this technique is efficient enough for use in practice.

## 1 Top-down motivation: End cargo cult parsing

"Cargo cult parsing" is a plague upon computing.[1] *Cargo cult parsing* refers to the use of "magic" regular expressions—often cut and pasted directly from Google search results—to parse languages which ought to be parsed with context-free grammars. Such parsing has two outcomes. In the first case, the programmer produces a parser that works "most of the time" because the underlying language is fundamentally *ir*regular. In the second case, some domain-specific language ends up with a mulish syntax, because the programmer has squeezed the language through the regular bottleneck. There are two reasons why regular expressions are so abused while context-free languages remain foresaken: (1) regular expression libraries are available in almost every language, while parsing libraries and toolkits are not, and (2) regular expressions are "WYSIWYG"—the language described is the language that gets matched—whereas parser-generators are WYSIWYGIYULR($k$)—"what you see is what you get if you understand LR($k$)." To end cargo-cult parsing, we need a new approach to parsing that:

1. handles arbitrary context-free grammars;
2. parses efficiently on average; and
3. can be implemented as a library with little effort.

The end goals of the three conditions are *simplicity*, *feasibility* and *ubiquity*. The "arbitrary context-free grammar" condition is necessary because programmers will be less inclined to use a tool that forces them to learn or think about LL/LR arcana. It is hard for compiler experts to imagine, but the constraints on LL/LR grammars are (far) beyond the grasp of the *average* programmer. [Of course, arbitrary context-free grammars bring ambiguity, which means the parser must be prepared to return a parse forest rather than a parse tree.]

---

[1] The term "cargo cult parsing" is due to Larry Wall, 19 June 2008, Google Tech Talk.

The "efficient parsing" condition is necessary because programmers will avoid tools branded as *inefficient* (however justly or unjustly this label has been applied). Specifically, a parser needs to have roughly linear behavior on average. Because ambiguous grammars may yield an exponential number of parse trees, parse trees must be produced lazily, and each parse tree should be paid for only if it is actually produced.

The "easily implemented" condition is perhaps most critical. It must be the case that a programmer could construct a general parsing toolkit if their language of choice doesn't yet have one. If this condition is met, it is reasonable to expect that proper parsing toolkits will eventually be available for every language. When proper parsing toolkits and libraries remain unavailable for a language, cargo cult parsing prevails.

This work introduces parsers based on the derivative of context-free languages and upon the derivative of parser combinators. Parsers based on derivatives meet all of the aforementioned requirements: they accept arbitrary grammars, they produce parse forests efficiently (and lazily), and they are easy to implement (less than 250 lines of Scala code for the complete library).[2] Derivative-based parsers also avoid the precompilation overhead of traditional parser generators; this cost is amortized (and memoised) across the parse itself. In addition, derivative-based parsers can be modified mid-parse, which makes it conceivable that a language could to modify its own syntax at compile- or run-time.

## 2 Bottom-up motivation: Generalizing the derivative

Brzozowski defined the derivative of regular expressions in 1964 [1]. This technique was lost to the "sands of time" until Owens, Reppy and Turon recently revived it to show that derivative-based lexer-generators were easier to write, more efficient and more flexible than the classical regex-to-NFA-to-DFA generators [15]. (Derivative-based lexers allow, for instance, both complement and intersection.)

Given the payoff for regular languages, it is natural to ask whether the derivative, and its benefits, extend to context-free languages, and transitively, to parsing. As it turns out, they do. We will show that context-free languages are closed under the derivative—they critical property needed for parsing. We will then show that context-free *parser combinators* are also closed under a generalization of the derivative. The net impact is that we will be able to write a derivative-based parser combinator library in under 250 lines of Scala, capable of producing a lazy parse forest for any context-free grammar.

---

[2] The second author on this paper, an undergraduate student, completed the implementation for Haskell in less than a week.

## 3 Contributions

1. An extension of the derivative to context-free languages.
2. A generalization of the derivative to parser combinators.
3. Two new methods for constructing a parse forest:
   - The first method uses two small-step machines; the first machine emits all possible parse strings as it executes; the second machine consumes the output of the first to generate a parse forest.
   - The second method redefines a parser combinator in terms of its derivative, yielding a simple recursive algorithm for generating lazy parse forests.
4. Implementations of derivative-based parsing combinators for Scala and Haskell:

   http://www.ucombinator.org/projects/parsing/

## 4 Preliminaries: Formal languages

A **formal language** $L \subseteq A^*$ is a set of sequences (or **words**) over an alphabet $A$. The set of all formal languages over the alphabet $A$ is the set $\mathbb{L}_A = \mathcal{P}(\mathcal{P}(A^*))$.

Rather than write words using sequence notation, we use juxtaposition, so that $a_1 \ldots a_n \equiv \langle a_1, \ldots, a_n \rangle$. When a word $w$ is used in a context that expects a language, the word is implicitly promoted to a singleton set, so that $w \equiv \{w\}$. The **empty language** is the empty set, $\emptyset$, while the **null string** $\epsilon$, is the string with no characters: $\epsilon = \langle \rangle$.

*Regular operations* Given two languages $L_1$, $L_2$, the **concatenation** of these two languages, denoted $L_1 \cdot L_2$, is all possible appendings of strings in the first set with those in the second set:

$$L_1 \cdot L_2 = \{w_1 w_2 : w_1 \in L_1, w_2 \in L_2\}.$$

Because languages are sets, they may of course be unioned ($L_1 \cup L_2$), intersected ($L_1 \cap L_2$), complemented ($\overline{L}$) and so on.

The $n$th **power of a language** $L$, denoted $L^n$, is that language concatenated with itself $n$ times:

$$L^n = \{w_1 \ldots w_n : w_i \in L \text{ for all } i\},$$

where the zeroth power of a language is the set containing only the null string:

$$L^0 = \{\epsilon\}.$$

The **closure**, or **Kleene star**, of a language, $L$, denoted $L^\star$, is the union of all powers of a language, starting from 0:

$$L^\star = \bigcup_{i \geq 0}^{\infty} L^i.$$

*Context-free languages* A language is **context-free** if it may be expressed as a set of recursive equations over union and concatenation. For example, the set *List* is the language of non-empty sequences of the character x:

$$List = (List \cdot \{\texttt{x}\}) \cup \{\texttt{x}\}$$

where the implicit interpretation of a set of recursive language equations over monotonic functions $f_1, \ldots, f_n$:

$$X_1 = f_1(X_1, \ldots, X_n)$$
$$\vdots \qquad \vdots$$
$$X_n = f_n(X_1, \ldots, X_n)$$

is the *least* fixed point of the function $F$:

$$F(X_1, \ldots, X_n) = (f_1(X_1, \ldots, X_n),$$
$$\vdots$$
$$f_n(X_1, \ldots, X_n))$$

Context-free languages may also be defined with context-free grammars. A **context-free grammar** is a 4-tuple $(A, N, R, n_0)$. The alphabet $A$ contains terminal symbols. The set $N$ contains nonterminal symbols. (The set $S = A \cup N$ contains terminal and nonterminal symbols.) The set $R \subseteq N \times S^*$ contains rules of the form $n \to s_1 \ldots s_n$. And, the nonterminal $n_0$ is the initial nonterminal.

The language accepted by a grammar $G$ is the set $\mathcal{L}(G)$:

$$\frac{s_1 \ldots n \ldots s_m \in \mathcal{L}(A, N, R, n_0) \qquad n \to w \in R}{s_1 \ldots w \ldots s_m \in \mathcal{L}(A, N, R, n_0),}$$

and by default, the start symbol is in the language:

$$n_0 \in \mathcal{L}(A, N, R, n_0).$$

*Properties of languages* A language $L$ is **nullable** if it accepts the null string—if $\epsilon \in L$. The nullability function $\delta : \mathbb{L}_A \to \{\{\epsilon\}, \emptyset\}$ determines whether a language is nullable by returning $\{\epsilon\}$ if it is, and $\emptyset$ if not:

$$\delta(L) = \begin{cases} \{\epsilon\} & \epsilon \in L \\ \emptyset & \epsilon \notin L. \end{cases}$$

For regular languages, nullability is easily computable, and for context-free grammars, the nullability of all nonterminal symbols can be computed simultaneously with a straightforward fixed-point computation.

The **first set** of a language is the set of characters which may appear first in strings in that language:

$$first(L) = \{c : cw \in L \text{ for some } w\}.$$

The first sets of all nonterminals can be co-computed during the same fixed-point computation used to determine nullability.

## 5 The derivative of a formal language

In this section, we review the derivative of formal languages, as defined by Brzozowski [1]. We will examine properties of the derivative, including the properties that prove its closure under regular languages. We will also show how the derivative may be used to prove membership in a formal language. The next section will construct the derivative for context-free languages.

The **derivative of a formal language** $L$ with respect to a character $c$, denoted $D_c(L)$, is the set of tails for all strings beginning with the character $c$:

$$D_c(L) = \{w : cw \in L\}.$$

For example, $D_\mathtt{f}\{\mathtt{foo}, \mathtt{bar}, \mathtt{frak}\} = \{\mathtt{oo}, \mathtt{rak}\}$. For regular expressions, the derivative's chief utility lies in its ability to determine membership in a regular language without NFA or DFA conversion. In fact, Brzozowski's result implies that for *any* class of decidable languages closed under the derivative, derivatives may be used to determine membership. For a string of length $n$, it takes $n$ applications of the following (bidirectional) inference rule to determine membership in a language:

$$\frac{cw \in L}{w \in D_c(L).}$$

In short, to check if $w \in L$, take tails of the string and derivatives of the language until the string is the null string, and then check whether the final language accepts the null string.

Brzozowski showed that the derivative is closed under regular operations, and that the derivative is computable with straightforward structural recursion. For both the null language and the empty language, the derivative is empty:

$$D_c\{\epsilon\} = \emptyset$$
$$D_c(\emptyset) = \emptyset.$$

For single-character languages, the derivative is either the empty language or the null language:

$$D_c\{c\} = \{\epsilon\}$$
$$D_c\{c'\} = \emptyset \text{ if } c \neq c'.$$

For the concatenation of two languages, the derivative must account for the possibility that the first language is nullable:

$$D_c(L_1 \cdot L_2) = D_c(L_1) \cdot L_2 \cup \delta(L_1) \cdot D_c(L_2).$$

The term $\delta(L_1)$ acts as a canceling or enabling term, since:

$$\{\epsilon\} \cdot L = L$$
$$\emptyset \cdot L = \emptyset.$$

The derivative of a union is the union of the derivative:

$$D_c(L_1 \cup L_2) = D_c(L_1) \cup D_c(L_2).$$

The derivative of the complement is the complement of the derivative:

$$D_c(\overline{L}) = \overline{D_c(L)}.$$

In fact, for any standard set operation $\otimes \in \{\cup, \cap, -, \ldots\}$:

$$D_c(L_1 \otimes L_2) = D_c(L_1) \otimes D_c(L_2).$$

## 6 The derivative of a context-free language

In this section, we'll show that it is possible to construct the derivative of a context-free language by transforming a context-free grammar for that language. We will also discuss optimizations to simplify the computation of the derivative. The subsequent section will show how the derivative may be harnessed to produce the parse forest of a string, instead of just determining membership.

Because of the recursion in context-free grammars, the straightforward approach to computing the derivative does not work. For instance, consider the derivative of the *List* language in the preliminaries with respect to x:

$$\begin{aligned} D_x(List) &= D_x((List \cdot \{x\}) \cup \{x\}) \\ &= D_x(List \cdot \{x\}) \cup \{\epsilon\} \\ &= D_x(List) \cdot \{x\} \cup \{\epsilon\} \\ &= (D_x(List) \cdot \{x\} \cup \{\epsilon\}) \cdot \{x\} \cup \{\epsilon\} \\ &\ldots \end{aligned}$$

That is, when we attempt interpret the derivative computationally, we can easily find infinite recursion with no base case. (Of course, this is not a problem for programming languages that allow laziness, a fact which we will exploit in our implementation.)

Fortunately, we can construct the derivative based on the context-free grammar representation to avoid ill-foundedness. Given a context-free grammar $G = (A, N, R, n_0)$, the derivative of the language $\mathcal{L}(G)$ with respect to the character $c$ is the language $D_c(\mathcal{L}(G))$:

$$D_c(\mathcal{L}(G)) = \mathcal{L}(A, N', R', n'_0), \text{ where}$$

– the derivative of a single alphabet character $a \in A$ is the null string terminal or the empty nonterminal:

$$D_c(a) = \begin{cases} \epsilon & c = a \\ \emptyset & c \neq a \end{cases} ; \text{ and}$$

- for each nonterminal $n \in N$, we define a new, distinct nonterminal, $D_c(n) \notin N$, so that:[3]
$$N' = N \cup \{D_c(n) : n \in N\}\text{; and}$$
- for each rule $(n \to s_1 \ldots s_m)$, if the sequence $s_1 \ldots s_i$ is nullable, then:
$$[D_c(n) \to D_c(s_{i+1})s_{i+2} \ldots s_m] \in R',$$
and if $m = 0$, then:
$$D_c(n) \to \emptyset,$$
and:
$$(n \to s_1 \ldots s_m) \in R'\text{; and}$$
- $n'_0 = D_c(n_0)$.

**Theorem 1.** *Context-free languages are closed under the derivative.*

*Proof.* By the prior construction.

### 6.1 Optimizing the derivative

The aforementioned process for computing the derivative is simple and correct, but clearly inefficient. It results in a quadratic blowup in rule size with every derivative. First, it computes the derivative of every rule, whether or not it's necessary. In pratice, a derivative rule should be inserted for a symbol $n$ only once its derivative appears in another rule. Second, if the empty nonterminal appears in a rule, $n \to w\emptyset w'$, then the rule should be replaced with $n \to \emptyset$. If the only rule for a nonterminal is the empty rule, then instances of that nonterminal in other rules should be replaced with the empty set too. The empty nonterminal arises frequently with derivatives, and if this reduction is applied aggressively, the grammar will remain a manageable size.

## 7 Approach 1: Parsing with derivatives of parser combinators

In this section, we review nondeterministic parser combinators. We then show how to generalize the concept of the derivative to these parser combinators, and how to use the derivative to compute a parse forest in a straightforward fashion.

Popular as of late, parser combinators are an elegant mechanism for embedding parsers inside an existing programming language. A type-$X$ parser combinator over an alphabet $A$ is a function, $\tau_X$, which consumes a sequence of characters and produces all possible parses of (non-strict) prefixes of that sequence. Formally, a type-$X$ parser combinator consumes a sequence of characters; it produces a set of pairs; each pair contains a value from the set $X$ and the remainder of the input after having parsed the value:

$$\tau_X \in \mathcal{T}_X = A^* \to \mathcal{P}(X \times A^*).$$

---

[3] Borrowing a trick from ordinal theory, we could define $D_c(n) = \{n\}$.

*Example 1.* If we had a parser combinator over non-empty sequences of the character z, $\tau_{\mathsf{z}^*} : A^* \to \{\mathsf{z}\}^*$, then:

$$\tau_{\{\mathsf{z}\}^*}(\mathsf{zzz}) = \{(\mathsf{z},\mathsf{zz}), (\mathsf{zz},\mathsf{z}), (\mathsf{zzz},\epsilon)\}.$$

Parser combinators must meet two conditions: (1) monotonicity and (2) contractiveness. A parser combinator is monotonic iff

$$w \sqsubseteq w' \text{ implies } \tau_X(w) \sqsubseteq \tau_X(w').$$

Over words, the partial order ($\sqsubseteq$) is "is a prefix of," and over pairs:

$$(x, w) \sqsubseteq (x', w') \text{ iff } x = x' \text{ and } w \sqsubseteq w',$$

In addition, parser combinators must be contractive; every string in the output must be a (non-strict) suffix of the input string:

$$(x, w') \in \tau_X(w) \text{ implies } w' \text{ is a suffix of } w.$$

When we want to parse an entire string, with no remainder, we can compute the **full combinator** of $\tau_X$, denoted $\lfloor \tau_X \rfloor : A^* \to \mathcal{P}(X)$:

$$\lfloor \tau_X \rfloor(w) = \{x : (x, \epsilon) \in \tau_X(w)\}.$$

That is, the full combinator discards any parse that didn't consume the entire input. Parser combinators do describe a formal language:

$$\mathcal{L}(\tau_X) = \{w : \tau_X(w) \text{ contains } (x, \epsilon) \text{ for some } x \in X\}.$$

### 7.1 Operations or combinators

A parser combinator is a black box. What makes them useful is the ability the combine them to form new combinators. There are operations on combinators that are analogous to the operations on formal languages. For instance, combinators may be concatenated, so that $\tau_X \cdot \tau_Y \in \mathcal{T}_{X \times Y}$:

$$\tau_X \cdot \tau_Y = \lambda w. \{((x, y), w'') : (x, w') \in \tau_X(w), (y, w'') \in \tau_Y(w')\}.$$

In other words, the concatenation runs the first parser combinator on the input to produce parse nodes of type $X$, and then it runs the second combinator on the remainders of the input to produce parse nodes of type $Y$. In addition to concatenation, combinators may also be unioned, so that $\tau_X \cup \tau'_X \in \mathcal{T}_X$:

$$\tau_X \cup \tau'_X = \lambda w. \tau_X(w) \cup \tau'_X(w).$$

It is convenient to turn a string into a parser combinator for itself:

$$w \equiv \lambda w'. \begin{cases} \{(w, w'')\} & w' = ww'' \\ \emptyset & \text{otherwise} \end{cases} \in \mathcal{T}_{A^*}.$$

There is an empty combinator as well:

$$\emptyset \equiv \lambda w.\emptyset.$$

Some operations on combinators, such as reduction, have no analogy in the formal language space. The **reduction** of combinator $\tau_X$ with respect to the function $f : X \to Y$, denoted $\tau_X \to f$, maps this function over the output of a combinator, so that $\tau_X \to f \in \mathcal{T}_Y$:

$$\tau_X \to f = \lambda w. \{(f(x), w') : (x, w') \in \tau_X(w)\} \in \mathcal{T}_Y$$

*Example 2.* Parser combinators make it easy to define the syntax and semantics of simple languages together. For example, for the simple expression language, where $A = \{\texttt{+},\texttt{*},\texttt{(},\texttt{)},\texttt{0},\texttt{1}\}$, the meaning of an input expression $w \in A^*$ is contained within $\tau_\mathbb{N}(w)$:

$$\begin{aligned}
\tau_\mathbb{N} &= \tau'_\mathbb{N} \\
&\cup \tau'_\mathbb{N} \cdot \texttt{+} \cdot \tau_\mathbb{N} &&\to \lambda(n_1, \texttt{+}, n_2).n_1 + n_2 \\
\\
\tau'_\mathbb{N} &= \tau''_\mathbb{N} \\
&\cup \tau''_\mathbb{N} \cdot \texttt{*} \cdot \tau'_\mathbb{N} &&\to \lambda(n_1, \texttt{*}, n_2).n_1 \times n_2 \\
\\
\tau''_\mathbb{N} &= \texttt{0} &&\to 0 \\
&\cup \texttt{1} &&\to 1 \\
&\cup \texttt{(} \cdot \tau_\mathbb{N} \cdot \texttt{)} &&\to \lambda(\texttt{(}, n, \texttt{)}).n
\end{aligned}$$

For example, $(4, \epsilon) \in \tau_\mathbb{N}(\texttt{(1+1)*(1+1)})$.

### 7.2 Derivatives of parser combinators

Remarkably, it is possible to generalize derivatives to parser combinators. Even more remarkably, we can use identities involving the derivative to *implement* parser combinators.

The **derivative of a parser combinator**, $D_c : \mathcal{T}_X \to \mathcal{T}_X$ returns a new parser combinator; when applied, the new combinator produces what the original combinator would return if the specified character were forcibly consumed first:

$$D_c(\tau_X) = \lambda w.\tau_X(cw) - (\lfloor \tau_X \rfloor(\epsilon) \times \{cw\}).$$

The difference operation in the derivative discards parse trees that can be generated without consuming the specified character—all those parse trees which consumed no input. Without the difference operation, it is possible that a derivative might not contract the input—that $(x, cw) \in D_c(\tau_X)(w)$ could happen.

When generating parse trees with derivatives of combinators, we can derive a straightforward identity from the definition of the derivative (and the monotonicity of combinators) that leads to a naturally recursive implementation of a

parser combinator:

$$D_c(\tau_X) = \lambda w.\tau_X(cw) - (\lfloor \tau_X \rfloor(\epsilon) \times \{cw\})$$
$$\text{iff } D_c(\tau_X)(w) = \tau_X(cw) - (\lfloor \tau_X \rfloor(\epsilon) \times \{cw\})$$
$$\text{iff } D_c(\tau_X)(w) \cup (\lfloor \tau_X \rfloor(\epsilon) \times \{cw\}) = \tau_X(cw).$$

Or, from the combinator's perspective:

$$\tau_X(cw) = (D_c(\tau_X))(w) \cup (\lfloor \tau_X \rfloor(\epsilon) \times \{cw\}). \tag{1}$$

A useful optimization exploits the fact that a full combinator, which is not interested in partial results, obeys a simpler recursive identity:

$$\lfloor \tau_X \rfloor(cw) = \lfloor D_c(\tau_X) \rfloor(w). \tag{2}$$

This equation is almost the entire parsing algorithm by itself. All it lacks is base case—a method for parsing the empty string.

### 7.3 Derivatives under operations

To make this implementation strategy possible, we need to show that it is possible to compute the derivative of a parser combinator composed of standard operations. If a combinator is the concatenation of two combinators, then:

$$D_c(\tau_A \cdot \tau_B) = \begin{cases} D_c(\tau_A) \cdot \tau_B & \epsilon \notin \mathcal{L}(\tau_A) \\ D_c(\tau_A) \cdot \tau_B \cup (\epsilon \to \lambda \epsilon. \lfloor \tau_A \rfloor(\epsilon)) \cdot D_c(\tau_B) & \text{otherwise.} \end{cases} \tag{3}$$

The derivative distributes across union:

$$D_c(\tau_A \cup \tau_B) = D_c(\tau_A) \cup D_c(\tau_B). \tag{4}$$

On single-character combinators, the derivative removes a character, or obliterates the combinator:

$$D_c(c) = \epsilon \to \lambda w. \{(c, w)\}$$
$$D_c(c') = \emptyset \text{ if } c \neq c'.$$

The derivative of either the null string combinator or the empty combinator is the empty combinator:

$$D_c(\epsilon) = \emptyset$$
$$D_c(\emptyset) = \emptyset.$$

The derivative of a reduction is the reduction of the derivative:

$$D_c(\tau_X \to f) = D_c(\tau_X) \to f.$$

### 7.4 Parsing with the derivative of parser combinators

To compute the derivative of concatenation, it is necessary to be able parse the empty string. Equation 2 is the inductive step, but at the very end of the algorithm, the combinator must parse the null string to yield the parse forest. Fortunately, the result of a parser applied to the null string is easy to compute with a least-fixed-point computation over the following relations:

$$\lfloor \emptyset \rfloor(\epsilon) = \emptyset \tag{5}$$
$$\lfloor \epsilon \rfloor(\epsilon) = \{\epsilon\} \tag{6}$$
$$\lfloor c \rfloor(\epsilon) = \emptyset \tag{7}$$
$$\lfloor \tau_X \cdot \tau_Y \rfloor(\epsilon) \supseteq \lfloor \tau_X \rfloor(\epsilon) \times \lfloor \tau_Y \rfloor(\epsilon) \tag{8}$$
$$\lfloor \tau_X \cup \tau_X' \rfloor(\epsilon) \supseteq \lfloor \tau_X \rfloor(\epsilon) \tag{9}$$
$$\lfloor \tau_X \cup \tau_X' \rfloor(\epsilon) \supseteq \lfloor \tau_X' \rfloor(\epsilon) \tag{10}$$
$$\lfloor \tau_X \to f \rfloor(\epsilon) \supseteq f.(\lfloor \tau_X \rfloor(\epsilon)). \tag{11}$$

This provides the base case of the recursive parsing method.

In summary, to convert a string into a set of parse trees, the string is consumed with the derivative, character-by-character, until the empty string is reached, at which point, a straightforward fixed point computation computes the trees that can arise from parsing the empty string. In practice, the cost of this fixed point computation may be amortized over the lifetime of the parse by computing it for sub-languages as they are derived.

### 7.5 Implementing parser combinators with laziness

In an implementation, all combinator arguments to either concatenation, union or reduction should be computed by need; that is, their computation should be delayed until necessary, and then the result should be cached. Computing by need (or at least lazily) prevents an implementation of the derivative from diverging on recursive languages such as:

$$\begin{aligned} X &= \mathbf{x} & &\to \lambda x.x \\ L &= L \cdot X & &\to \lambda(\boldsymbol{x}, x).\boldsymbol{x} \mathbin{+\!\!+} \langle x \rangle \\ &\cup\ X & &\to \lambda x.\langle x \rangle. \end{aligned}$$

In addition, the result of a parser combinator should be implemented as a lazy stream rather than an actual set, since parser combinators may produce an infinite number of results.

## 8 Approach 2: Parsing context-free languages with derivatives

For regular languages, the goal is determining membership. For context-free languages, the goal is to construct parse trees. To construct all possible parse trees for a string, we define a nondeterministic small-step parsing machine that will emit all possible parse *strings* as it executes. A second small-step machine will consume these parse strings and convert them into parse trees.

This approach is particularly well suited to implementing parsing libraries in *untyped* languages. This approach is not well suited to typed languages, since a literal implementation of the parsing machine must use a generic parse-tree type, or else violate type-safety with casts for reductions. For typed languages, the parser combinator approach of the next section yields a simpler (type-safe) implementation.

The goal of a parser is to produce a parse tree. For a context-free grammar $G = (A, N, R, n_0)$, we can describe a generic parse tree as a member of the recursive structure $T$:

$$t \in T = A + (R \times T^*).$$

In a parse tree, the leaf nodes are terminals, internal nodes are tagged with rules and children are the right-hand-side of a rule.

*Example 3.* For the grammar describing the language of balanced parentheses:

$$B \to (B)B$$
$$B \to \epsilon$$

the parse tree is:

$$
\begin{array}{c}
B \\
\overparen{(\ B\ )\ B} \\
|\quad\ | \\
\epsilon\quad \epsilon
\end{array}
$$

or formally, $(B \to (B)B, \langle (, (B \to \epsilon, \langle\rangle), ), (B \to \epsilon, \langle\rangle)\rangle)$.

A parse string is a linear encoding of a parse tree containing both nonterminal markers (written as a bra $\langle n|$) and reduction markers (written as a ket $|r\rangle$). A bra character denotes the start of an internal node, while the enclosing ket determines the end of an internal node, and with which rule to reduce.

*Example 4.* The parse string encoding of the tree from the prior example is
$\langle B|\ (\ \langle B|B \to \epsilon\rangle\ )\ \langle B|B \to \epsilon\rangle\ |B \to (B)B\rangle$.

To create parse strings over the grammar $G = (A, N, R, n_0)$, we first construct a new grammar $G' = (A', N, R', n_0)$, over parse strings of this grammar:

$$A' = A \cup \{\langle n| : n \in N\} \cup \{|r\rangle : r \in R\},$$

and for every $r \in R$, if $r = (n \to s_1 \ldots s_m)$, then

$$(n \to \langle n|s_1 \ldots s_m|r\rangle) \in R'.$$

Next, we construct a nondeterministic, small-step state machine where transitions are labeled with parse-string characters. Along any path this machine takes, it will emit—character-by-character—parse strings. The state of this machine ($\Sigma$) is a context-free language paired with an input string:

$$\varsigma \in \Sigma = \mathbb{L} \times A^*.$$

The semantics of the transition relation $(\Rightarrow) \subseteq \Sigma \times A' \times \Sigma$ requires one rule for producing a character:
$$(L, cw) \Rightarrow^c (D_c(L), w),$$

a second rule for producing bras:

$$\frac{\langle n| \in \mathit{first}(L)}{(L, w) \Rightarrow^{\langle n|} (L, w),}$$

and a third (and final) rule for producing kets:

$$\frac{|r\rangle \in \mathit{first}(L)}{(L, w) \Rightarrow^{|r\rangle} (L, w).}$$

We can build a second machine—a parsing machine—that consumes the output of the first machine to produce a parse tree. This parsing machine maintains a parsing stack whose elements are parse trees. Thus, a parsing machine state, $\psi \in \Psi$, is a parse-string machine state paired with this stack:

$$\psi \in \Psi = \Sigma \times T^*.$$

The transition relation has three rules as well. If the parse-string machine emits a character $c \in A$, the parsing machine pushes it on the stack:

$$\frac{\varsigma \Rightarrow^c \varsigma'}{(\varsigma, t) \twoheadrightarrow (\varsigma', c : t),}$$

and if the parse-string machine emits a non-terminal, it pushes that as well:

$$\frac{\varsigma \Rightarrow^{\langle n|} \varsigma'}{(\varsigma, t) \twoheadrightarrow (\varsigma', \langle n| : t),}$$

but if the parse-string machine emits a rule, the parsing machine reduces:[4]

$$\frac{\varsigma \Rightarrow^{|n \to s_1 \ldots s_m\rangle} \varsigma'}{(\varsigma, \langle t_m, \ldots, t_1\rangle \mathbin{+\!\!+} t') \twoheadrightarrow (\varsigma', (n \to s_1 \ldots s_m, \langle t_1, \ldots, t_m\rangle) : t').}$$

The parsing machine demonstrates that, even without reduction rules associated with the grammar, we can use derivatives to construct a concrete syntax tree.

---

[4] The symbol $\mathbin{+\!\!+}$ denotes sequence append.

## 9  Implementation of parser combinators

In this section, we discuss our experiences with an implementation of derivative-based parser combinators in Scala. Implementations of the combinator library for both Scala and Haskell are publicly available:

> http://www.ucombinator.org/projects/parsing/

Both implementations make pervasive use of laziness and caching to prevent infinite recursion during both the specification of grammars and the computation of derivatives.

In the Scala implementation parsers descend from the class `Parser[T,A]`:

```
abstract class Parser[T,A] {
 def derive (t : T) : Parser[T,A] ;
 def parseFull (input : Stream[T]) : Stream[A] ;
 def parse (input : Stream[T]) : Stream[(A,Stream[T])] ;
 ...
 def parseNull  : Stream[A]  ;
 def isNullable : Boolean ;
 def isEmpty    : Boolean ;
 ...
 def ||   (b : => Parser[T,A]) : Parser[T,A] ;
 def ~[B] (b : => Parser[T,B]) : Parser[T,~[A,B]] ;
 def *                         : Parser[T,List[A]] ;
 def ==>[B] (f : A => B)       : Parser[T,B] ;
}
```

A `Parser[T,A]` object represents a parser combinator which consumes streams over type `T` to produce a *lazy* `Stream` of parse trees of type `A`. The `derive` method computes the derivative of the combinator with respect to a token `t` (and caches the result); subclasses of `Parser[T,A]` have to implement this method. The `parseFull` method produces all possible parse trees which fully consume the input; its implementation follows Equation 2:

```
def parseFull (in : Stream[T]) : Stream[A] =
 if (in.isEmpty) this.parseNull
 else            this.derive(in.head).parseFull(in.tail)
```

The method `parseNull` uses a fixed-point computation (Equations 5-11) to produce the parse forest with respect to this combinator and the null string. The method `parse` follows Equation 1:

```
def parse (in : Stream[T]) : Stream[(A,Stream[T])] =
 if (in.isEmpty)
  this.parseNull map (_,Stream.empty)
 else
  this.derive(in.head).parse(in.tail) append
  for (a <- this.parseNull yield (a,in))
```

The properties `isEmpty` and `isNullable` are co-computed along with `parseNull` using fixed points. The methods `~`, `||`, `*`, `==>` are syntactic sugar for concatenation, union, closure and reduction respectively.

There are four main combinators, each represented as a subclass of `Parser`:

```
class Con[T,A,B] (a : => Parser[T,A], b : => Parser[T,B])
  extends Parser[T,~[A,B]]

class Alt[T,A]   (a : => Parser[T,A], b : => Parser[T,A])
  extends Parser[T,A]

class Rep[T,A]   (p : => Parser[T,A])
  extends Parser[T,List[A]]

class Red[T,A,B] (p : => Parser[T,A], f : A => B)
  extends Parser[T,B]
```

It is important to note that all parser parameters are passed *by name*. And, each is accessed through a `lazy` field that caches the result upon first reference. This laziness makes it possible to compute the derivative of recursive grammars: since the computation of subsequent derivatives is *immediately* suspended.

The derivative of catenation follows Equation 3, short-circuiting if the first component is nullable:

```
protected def internalDerive (t : T) : Parser[T,~[A,B]] =
  if (first.isNullable)
    new Alt(new Con(first.derive(t), second),
            new Con(new ε[T,A](first.parseNull),
                    second.derive(t)))
  else
    new Con(first.derive(t), second)
```

The derivative of union follows Equation 4, but short-circuits if it finds either arm is empty:

```
protected def internalDerive (t : T) : Parser[T,A] =
  if      (choice1.isEmpty) choice2.derive(t)
  else if (choice2.isEmpty) choice1.derive(t)
  else    /*  otherwise  */ new Alt(choice1.derive(t),
                                    choice2.derive(t))
```

There is a class for parsing an individual token (`T`), a class for the null string ($\epsilon$), and a class for the empty language ($\emptyset$):

```
class T[T]    (t : T)             extends Parser[T,T]
class ε[T,A]  (g : => Stream[A])  extends Parser[T,A]
class ∅[T,A]                      extends Parser[T,A]
```

The derivative for each of these is also straightforward, and directly follows their respective equations.

### 9.1 Experience with the implementation; the need for closure

Our original discussion of parser combinators did not include the closure operation, yet we have included it (as `Rep`) in our implementation. We introduced the closure operations for three reasons:

1. Regular expressions allow closure, which makes the operation familiar.
2. Many grammars feature sequences, so closures simplify specifications.
3. Parsing sequences can be heavily optimized, leading to performance gains.

Our original implementation did not contain the closure operation, and when we tested it by parsing large, randomly generated S-Expressions, parsing took time exponential in the size of the input; it took seconds to parse a hundred tokens, but over an hour to parse a thousands tokens. Using right-recursive lists reduced the performance penalty to minutes to parse thousands of tokens, but this was still unacceptably high.

By introducing the special closure construct, we achieved roughly linear scaling by hand-optimizing the implementation of the `parse`, `parseFull` and `derive` methods. The `derive` method follows the expected equation:

$$D_c(\tau_X^\star) = (D_c(\tau_X) \cdot \tau_X^\star) \to \lambda(x', \boldsymbol{x}).x' : \boldsymbol{x}.$$

The hand-optimization of the `parse` methods has them seek the longest possible parse first, which turns out to be precisely the behavior that programmers expect out of these constructs—in the same way they expect regular expressions to produce the longest match. The following table shows parsing time on randomly generated S-Expressions of increasing orders of magnitude in terms

| Number of tokens | Parsing time |
|---|---|
| 4,944 | 10 ms |
| 42,346 | 43 m |
| 390,553 | 326 ms |
| 5,049,213 | 3.9 s |
| 22,155,402 | 17.1 s |

These results suggest that derivative-based parser combinators scale well enough to be useful in practice.

## 10 Related work

Grammars and formal languages date to the earliest days of computer science [2]. Parsing is almost as ancient. Even the derivative of a regular expression is decades old [1]. Given all that has been done, it is surpising to find something was still left undiscovered. Perhaps not surprisingly, derivative-based parsing defies the classical taxonomies. Historically, methods for parsing have been divided into two groups: top-down and bottom-up. Derivative-based parsing does not fit cleanly into either category.

Top-down methods attempt to construct the parse tree by starting from the root, and predicting (with look-ahead) their way down to the leaves, backtracking whenever necessary, and typically caching results from failed attempts. Top-down methods include recursive descent parsing [19], LL($k$) parsing, and packrat/PEG parsing [13, 18]. Top-down methods differ from derivative-based methods in that, with the exception of Warth *et al.* [18], these methods have not been able to handle left-recursive grammars. Top-down methods, however, do lend themselves to combinator libraries, a strength which derivative-based techniques share. Philosophically, neither derivative-based method "feels" like a top-down parsing method.

Bottom-up methods compute the leaves of the parse tree first, and grow toward the root. Yet, derivative-based parsing differs from bottom-up methods too, including abstract interpretation [4, 5], operator-precedence parsing [11, 16], simple precedence parsing [8], bounded context parsing [12], SLR parsing [7], LALR parsing [6], LR($k$) parsing [14], GLR parsing [17], CYK parsing [10, 20, 3], and Earley parsing [9]. Derivative-based parsing shares full coverage of all context-free grammars with GLR, CYK and Earley. However, bottom-up methods tend to be poor choices for parser-combinator libraries, because they tend to precompile the grammar into a push-down machine. To they extent that derivative-based methods pre-compile, the cost of that compilation is amortized across the entire parsing process.

The parsing-machine approach to parse-tree construction has a bottom-up feel to it, and yet, when one examines the workings of the parse-string-*generating* machine, it still lacks a characteristic "bottom-up flavor." Derivative-based parser combinators are even less bottom-up than the parsing machine approach. If one examines the behavior of the derivative-based combinators as they execute, one can see that these combinators are not constructing parse trees either bottom-up or top-down; rather, these combinators are slowly transforming and unfolding the grammar itself into a parse tree; at any given state of the parse, the derived combinator is really part-forest, part-grammar.

Ultimately, derivative-based methods share the best properties of both words: they are easy to implement, and easy to implement as libraries (like top-down parsers), yet they are efficient and expressive (like bottom-up parsers).